# IMPROVED PRECISION MEASUREMENT OF THE CASIMIR FORCE USING GOLD SURFACES

*B.W. Harris, F. Chen and U. Mohideen\**
*Dept. of Physics, Univ. of California, Riverside, CA 92521.*

## ABSTRACT

We report an improved precision measurement of the Casimir force using metallic gold surfaces. The force is measured between a large gold coated sphere and flat plate using an Atomic Force Microscope. The use of gold surfaces removes some theoretical uncertainties in the interpretation of the measurement. The forces are also measured at smaller surface separations. The complete dielectric spectrum of the metal is used in the comparison of theory to the experiment. The average statistical precision remains at the same 1% of the forces measured at the closest separation. These results should lead to the development of stronger constraints on hypothetical forces.

The Casimir force [1,2] has its origin in the zero point electromagnetic vacuum fluctuations predicted by quantum electrodynamics. This zero point electromagnetic energy is given by $E = \sum_{n}^{\infty}(1/2)\hbar\omega_n$, where $\hbar\omega_n$ is the photon energy in each allowed photon mode $n$. If two perfectly reflecting metal plates, are held parallel then the alteration of the zero point energy by the metal boundaries leads to an attractive force between the plates called the Casimir force [1,2]. Lifshitz [3] generalized the force to any two infinite dielectric half-spaces as the force between fluctuating dipoles induced by the zero point electromagnetic fields and obtained the same result as Casimir for two perfectly reflecting (infinite conductivity) flat plates. The Casimir force has been demonstrated between two flat plates [4] and a large sphere and a flat plate [5,6] and its value shown to be in agreement with the theory to an average deviation of 1% [7-9]. For dielectric bodies the resulting force has been measured with reasonable agreement to the theory [10]. Theoretical treatments of the Casimir force have shown that it is a strong function of the boundary geometry and spectrum [11-13]. Experiments with periodically corrugated boundaries have also demonstrated the nontrivial boundary dependence of the Casimir force [14]. Here we report an improved precision measurement of the Casimir force between a metallized sphere of diameter 191.3 μm and a flat plate using an Atomic Force Microscope (AFM). The use of gold surfaces and the related experimental changes are the primary differences between the experiments reported here and the last version of the experiment [9]. In the previous experiments [7,9] Al surfaces were used due to their high reflectivity and ease of fabrication. However in order to prevent the effects of oxidation of the Al surfaces, a thin layer of sputtered Au/Pd was used on top of the Al surface. This thin Au/Pd coating was treated in a phenomenological manner in the earlier experiments [7-9]. Such a treatment complicates the theoretical analysis [15]. Thus it is necessary to use chemically inert materials such as gold for the measurement of the Casimir force which is reported here. The complete dielectric properties of Au is used in the theory. An important application of Casimir force measurements is to develop strong limits on hypothetical long range forces and light elementary particles such as those predicted by supersymmetric theories [16,17]. The use of gold surfaces with the higher densities should lead to large improvements in the calculated constraints of these hypothetical long range forces. The average precision defined on the rms deviation between experiment and theory remains at the same 1% of the forces measured at the closest separation. The measurement is consistent with the theoretical corrections calculated to date.

Given two parallel plates of unit area and infinite conductivity, separated by a distance $z$ the Casimir force is: $F(z) = -\frac{\pi^2 \hbar c}{240} \frac{1}{z^4}$. The force is a strong function of '$z$' and is measurable only for $z \leq 1$ μm. Experimentally it is hard to configure two parallel plates uniformly separated by distances on the order of a micron. So the preference is to replace one of the plates by a metal sphere of radius $R$ where $R \gg z$. For such a geometry the Casimir force is modified to [18]: $F_c^0(z) = \frac{-\pi^3}{360} R \frac{\hbar c}{z^3}$. This definition of the Casimir force holds only for hypothetical metals of infinite conductivity, and therefore a correction due to the finite conductivity of gold has to be applied. Such a correction can be accomplished through use of the Lifshitz theory [3,15,19]. For a metal with a dielectric constant $\varepsilon$ the force between a large sphere and flat plate is given by [3,15]:

$$F^0(z) = \frac{R\hbar}{2\pi c^2} \int_1^\infty \int_0^\infty p\, \xi^2 dp d\xi \left\{ \ln\left[1 - \frac{(K-p)^2}{(K+p)^2} e^{-\frac{2p\xi z}{c}}\right] + \ln\left[1 - \frac{(K-p\varepsilon)^2}{(K+p\varepsilon)^2} e^{-\frac{2p\xi z}{c}}\right]\right\} \quad (1)$$

where 'z' is the surface separation, $R$ is the sphere radius, $K = \sqrt{\varepsilon - 1 + p^2}$, $\varepsilon(i\xi) = 1 + \frac{2}{\pi}\int_0^\infty \frac{\omega \varepsilon''(\omega)}{\omega^2 + \xi^2} d\omega$ is the dielectric constant of gold and $\varepsilon''$ is its imaginary component. $\xi$ is the imaginary frequency given by $\omega = i\xi$. Here the complete $\varepsilon''$ extending from 0.125eV to 9919eV from Ref. [20] along with the Drude model below 0.125eV is used to calculate $\varepsilon(i\xi)$. In the Drude representation of the dielectric properties in terms of the imaginary frequency $\xi$, $\varepsilon(i\xi) = 1 + \frac{\omega_p^2}{\xi^2 + \gamma\xi}$, $\omega_p = 11.5$eV is the plasma frequency and $\gamma$ is the relaxation frequency corresponding to 50meV. These values of $\omega_p$ and $\gamma$ are obtained in the manner detailed in Ref. [21].

There are also corrections to the Casimir force resulting from the roughness of the metallic surfaces used. These corrections result from the stochastic changes in the surface separation [22]. Here the roughness of the metal surface is measured directly with the AFM. This leads to the complete Casimir force including roughness correction given by:

$F^r(z) = F^0(z)\left[1 + 6\left(\frac{A}{z}\right)^2\right]$ [22]. Here, $A$ is the mean roughness amplitude that is measured with the AFM. The roughness correction here is <<1% of the measured force. There are also corrections due to the finite temperature [23] given by:

$$F_c(z) = F^r(z)\left(1 + \frac{720}{\pi^2} f(\eta)\right), \tag{3}$$

where $f(\eta) = (\eta^3/2\pi)\zeta(3) - (\eta^4\pi^2/45)$, $\eta = 2\pi k_B Tz/hc = 0.131 \times 10^{-3} z$ nm$^{-1}$ for T= 300°K, $\zeta(3)$=1.202… is the Riemann zeta function and $k_B$ is the Boltzmann constant. The temperature correction is << 1% of the Casimir force for the surface separations reported here.

A schematic diagram of the experiment is shown in figure 1. The fabrication procedures had to be modified, given the different material properties of gold as compared to the aluminum coatings used previously in Ref. [7,9]. The 320μm long AFM cantilevers were first coated with about 200nm of aluminum to improve their thermal conductivity. This metal coating on the cantilever decreases the thermally induced noise when the AFM is operated in vacuum. Aluminum coatings are better, as applying thick gold coatings directly to these Silicon Nitride cantilevers led their curling due to the mismatch in the thermal expansion coefficients. Next polystyrene spheres were mounted on the tip of the metal coated cantilevers with Ag epoxy. A 1 cm diameter optically polished sapphire disk is used as the plate. The cantilever (with sphere) and plate were then coated with gold in an evaporator. The sphere diameter after the metal coating was measured using the Scanning Electron Microscope (SEM) to be 191.3±0.5μm. The rms roughness amplitude *A* of the gold surface on the plate was measured using an AFM to be 1.0±0.1nm. The thickness of the gold coating was measured using the AFM to be 96.2±0.7nm. Such a coating thickness is sufficient to reproduce the properties of an infinitely thick metal for the precisions reported here [15]. To reduce the development of contact potential differences between the sphere and the plate, great care was taken to follow identical procedures in making the electrical contacts. This is necessary given the large difference in the work function of aluminum and gold. The force is measured at a pressure below 30mTorr and at room temperature. The experiments were done on a floating optical table. As before, the vacuum system was mechanically damped and isolated to decrease the vibrations coupled to the AFM.

As shown in figure 1, a force between the sphere and plate causes the cantilever to flex. This flexing of the cantilever is detected by the deflection of the laser beam leading to a difference signal between photodiodes A and B. As in the previous measurements, this difference signal of the photodiodes was calibrated by means of an electrostatic force. The electrostatic force between the large sphere and the flat surface is given by [24]:

$$F = 2\pi\varepsilon_o (V_1 - V_2)^2 \sum_{n=1}^{\infty} \operatorname{csch} n\alpha (\coth \alpha - n \coth n\alpha) \quad . \quad (4)$$

Here '$V_1$' and '$V_2$' are voltages on the flat plate and sphere respectively. $\alpha = \cosh^{-1}\left(1 + \frac{z + z_0}{R}\right)$, where $R$ is the radius of the sphere, $z$ is distance between the surfaces, measured from contact and $z_o$ is the true average separation on contact of the two surfaces due to the stochastic roughness of the gold coating. For the measurement of the electrostatic force, the distance between the metallic surfaces was made >3μm (the distance is so chosen that the $z_o$ and the movement of the cantilever in response to the applied electrostatic force are negligible in comparison). Then various voltages $V_1$=+3V and $V_1$=-3V where applied to the plate while the sphere remained grounded. Given two polarities of the same voltage value for $V_1$, eq.4 was used to find the residual potential of the grounded sphere $V_2=3\pm3$ mV. This residual potential leads to forces which are <<1% of the Casimir forces at the closest separations reported here. Using this value of $V_2$ the photodiode difference signal of the AFM was calibrated from eq. 4. The movement of the piezoelectric tube on which the plate is mounted was calibrated by optical interferometry [25] and corrections (of order 1%) due to the piezo hysteresis were applied to the sphere-plate separations in all collected data.

To measure the Casimir force between the sphere and flat plate they are both grounded together with the AFM. The plate is then moved towards the sphere and the corresponding photodiode difference signal was measured (approach curve). The raw data from a scan is shown in Fig. 2. Region-1 is the flexing of the cantilever resulting from the continued extension of the piezo after contact of the two surfaces. In region-2 ($z_0$+400nm>surface separations>$z_o$nm) the Casimir force is the dominant characteristic far exceeding all systematic errors. The Casimir force measurement is repeated for 30 scans. The only systematic error associated with the Casimir force in these measurements is that due to the residual electrostatic force which is less than 0.1 % of the Casimir force at closest separation. For surface separations exceeding 400nm the experimental uncertainty in the force exceeds the value of the Casimir force. The surface separation on contact, $z_o$, is *a priori* unknown due to the roughness of the metal surface and is determined independently as described below. A small additional correction to the separation distance results from the deflection of the cantilever in response to the attractive Casimir force. As can be observed from the schematic in fig. 1, this leads to a decrease in the distance of separation of the two surfaces. This "deflection correction" modifies the separation distance between the two surfaces. This is given by: $z=z_o+z_{piezo}-F_{pd}*m$ , where $z$ is the correct separation between the two surfaces, $z_{piezo}$ is the distance moved by the plate due to the application of voltage applied to the piezo i.e the horizontal axis of figure 2 and $F_{pd}$ is the photodiode difference signal shown along the vertical axis in fig. 2. Here $m$ is deflection coefficient corresponding to the rate of change of seperation distance per unit photo-diode difference signal (from the cantilever deflection) and is determined independently as discussed below. The slope of the line in region-1 of the force curve shown in fig. 2 cannot be used to determine $m$ as the free movement of the sphere is prevented on contact of the two surfaces (due to the larger forces encountered here).

We use the electrostatic force between the sphere and flat plate to arrive at an independent measurement of the constant $m$ in the deflection correction and $z_o$ the average surface separation on contact of the two surfaces. This is done immediately following the Casimir force measurement without breaking the vacuum and no lateral movement of the surfaces. The flat plate is connected to a DC voltage supply while the sphere remains grounded. The applied voltage $V_1$ in eq. 4 is so chosen that the electrostatic force is much greater than the Casimir force. As can be observed from fig. 1, at the start of the force measurement, the plate and the sphere are separated by a fixed distance and the plate is moved towards the sphere in small steps with the help of the piezoelectric tube. When different voltages $V_1$ are applied to the plate, the point of contact between the plate and sphere varies corresponding to the different cantilever deflections. This is shown in fig. 3 for three different applied voltages 0.256, 0.202 and 0.154 V. The vertex in each curve identifies the contact point between sphere and plate. The deflection coefficient $m$ can be determined from the slope of the dashed line connecting the vertices. The slope corresponds to an average value of $m=8.9\pm0.3$ nm per unit photodiode difference signal. The separation distance is then corrected for this cantilever deflection. Next the surface separation on contact $z_o$ is determined from the same electrostatic force curves. The open squares in figure 4 represent the measured total force for an applied voltage of $V_1= 0.256$ V as a function of distance. The force results from a sum of the electrostatic force given by eq. 4 and the Casimir force of eq. 3. A best $\chi^2$ fit is done (shown as a solid line in the figure 4) to obtain the value of $z_o=31.7$ nm. The experiment is repeated for other voltages between 0.2-0.3 V leading to an average value of $z_o=32.7\pm0.8$ nm.

The average Casimir force measured from the 30 scans is shown as open squares in figure 5. The theoretical curve given by eq.3 is shown as a solid line. For clarity only 10% of the data points are shown in the figure. The error bars represent the standard deviation from the 30 scans at each data point. Due to the surface roughness, the averaging procedure introduces $\pm1$ nm uncertainty in the surface separation on contact of the two surfaces. The electrostatic force corresponding to the residual potential difference of $V_2=3$ mV has been subtracted from the measured Casimir force. As noted before this electrostatic force corresponds to less than 0.1% of the Casimir force at the closest separation.

A variety of statistical measures can be used to define the precision of the Casimir force measurement. A key point to note is that the Casimir force is generated for the whole range of separations and is compared to the theory with no adjustable parameters.

Thus we check the accuracy of the theoretical curve over the complete region between 62-350nm with $N=2583$ points (with an average of 30 measurements representing each point). Given that the experimental standard deviation around 62 nm is 19pN, the experimental uncertainty is $\leq \frac{19}{\sqrt{30}} = 3.5 pN$ leading to a precision which is better than 1% of the largest forces measured. If one wished to consider the rms deviation of the experiment ($F_{experiment}$) from the theory ($F_{theory}$) in eq.3, $\sigma = \sqrt{\frac{(F_{theory} - F_{experiment})^2}{N}} = 3.8$pN as a measure of the precision, it is also on the order of 1% of the forces measured at the closest separation. The uncertainties of 3.8pN measured here are larger than the 2 pN in previous measurements [7,9] due to the thinner gold coatings used which led to poor thermal conductivity of the cantilever. Thus experiments at cryogenic temperatures should substantially reduce the noise.

In conclusion, we have performed an improved precision measurement of the Casimir force between a large gold coated sphere and flat plate. As gold surfaces are chemically non-reactive, they do not require protective layers as used previously and thus the experimental results can be unambiguously compared to the theory. The complete dielectric properties of gold is used in the theory. The corrections due to the metal surface roughness and associated uncertainties in the contact separation have been substantially reduced. Also the electrostatic force due to the residual potential difference between the two surfaces has been lowered to negligible levels. The average precision defined on the basis of the rms deviation between experiment and theory remains at the same 1% of the forces measured at the closest separation. The measurement is consistent with the theoretical corrections calculated to date. The use of gold surfaces with their higher densities and the smaller separation distances at which the Casimir forces have been measured here, should lead to improvements in the calculated constraints of hypothetical long range forces such as those predicted by supersymmetric theories.

Discussions with G.L. Klimchitskaya, and V.M. Mostepanenko are acknowledged.

**FIGURE CAPTIONS**

Figure 1: Schematic diagram of the experimental setup. Application of voltage to the piezoelectric tube results in the movement of the plate towards the sphere. A force on the sphere leads to flexing of the cantilever.

Figure 2: The raw data of the force measured as a photodiode difference signal as a function of the distance moved by the plate.

Figure 3: The measured electrostatic force curves for three different voltages (a) 0.256 V (b) 0.202 V and (c) 0.154 V. The rate of change of separation distance per unit photodiode difference signal corresponding to the slope of the dashed line which connects the vertices yeilds the deflection coefficient *m*.

Figure 4: The measured electrostatic force for a applied voltage of 0.256 V to the plate is shown as open squares. For clarity only 10% of the points are shown in the figure. The best fit solid line shown leads to a $z_o$=31.7nm. The average of many voltages leads to $z_o$=32.7±0.8nm.

Figure 5: The measured average Casimir force as a function of plate-sphere separation is shown as squares. For clarity only 10% of the experimental points are shown in the figure. The error bars represent the standard deviation from 30 scans. The solid line is the theoretical Casimir force from eq. 3.

Figure 1, Harris, Feng and Mohideen

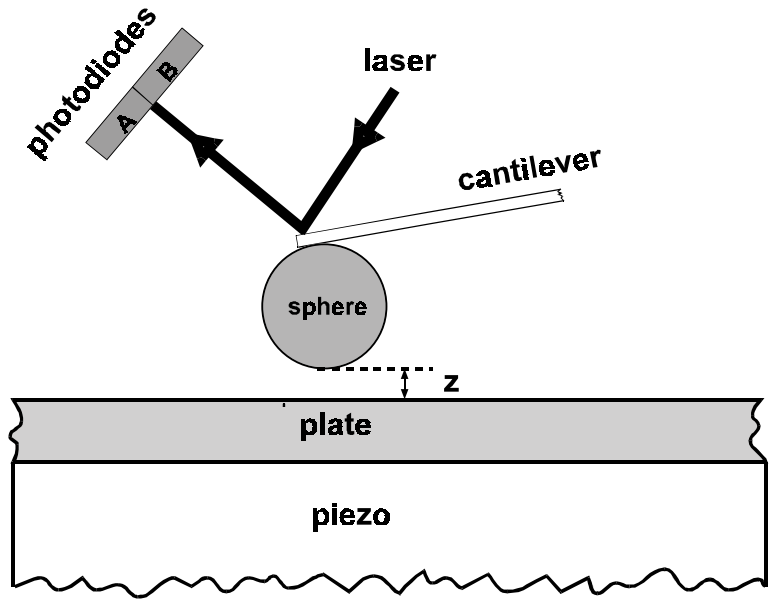

Figure 2, Harris, Feng and Mohideen

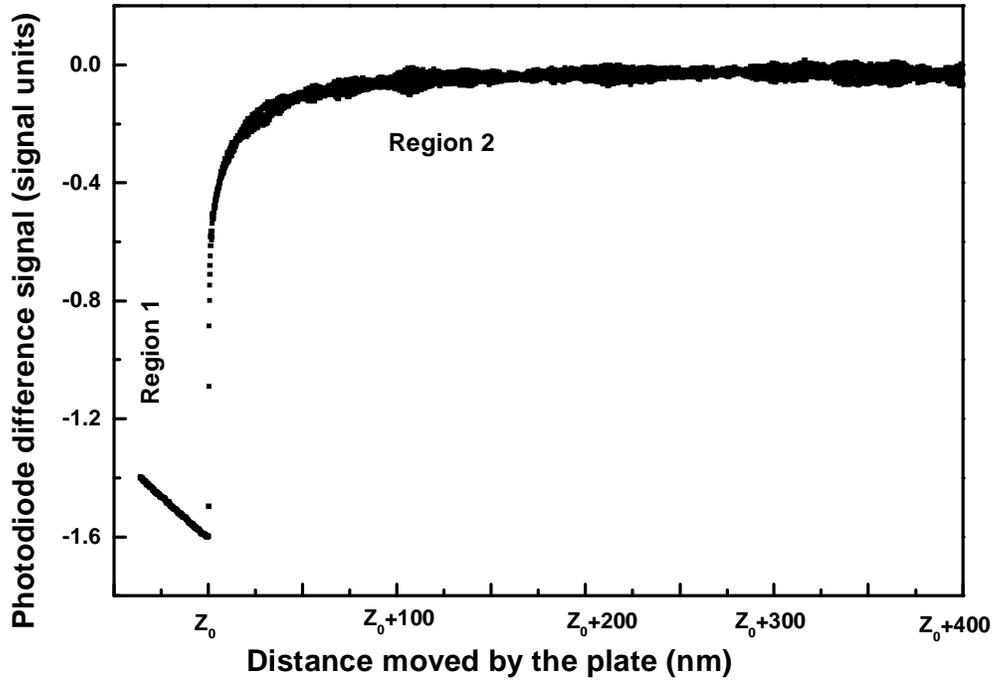

Figure 3, Harris, Feng and Mohideen

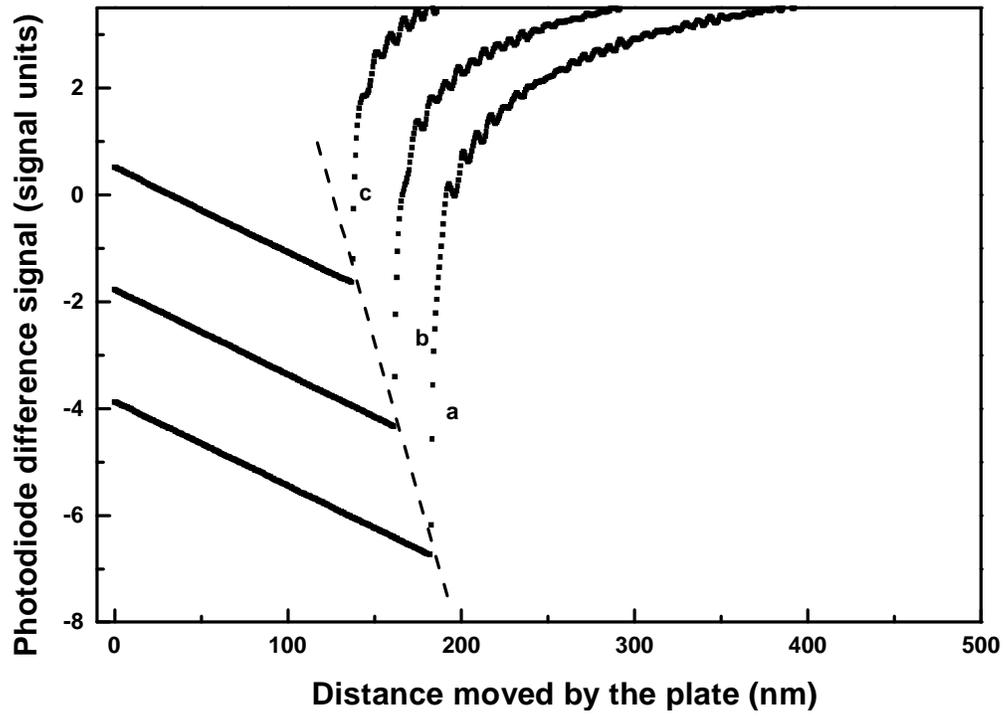

Figure 4  Harris, Feng and Mohideen

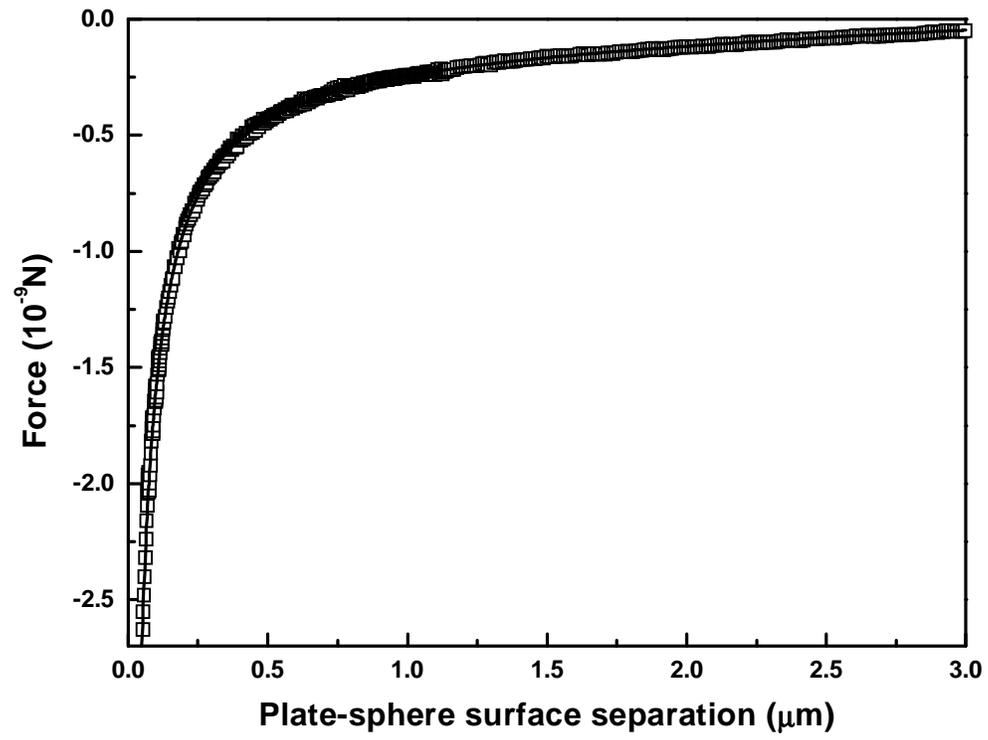

Figure 5: Harris, Feng and Mohideen

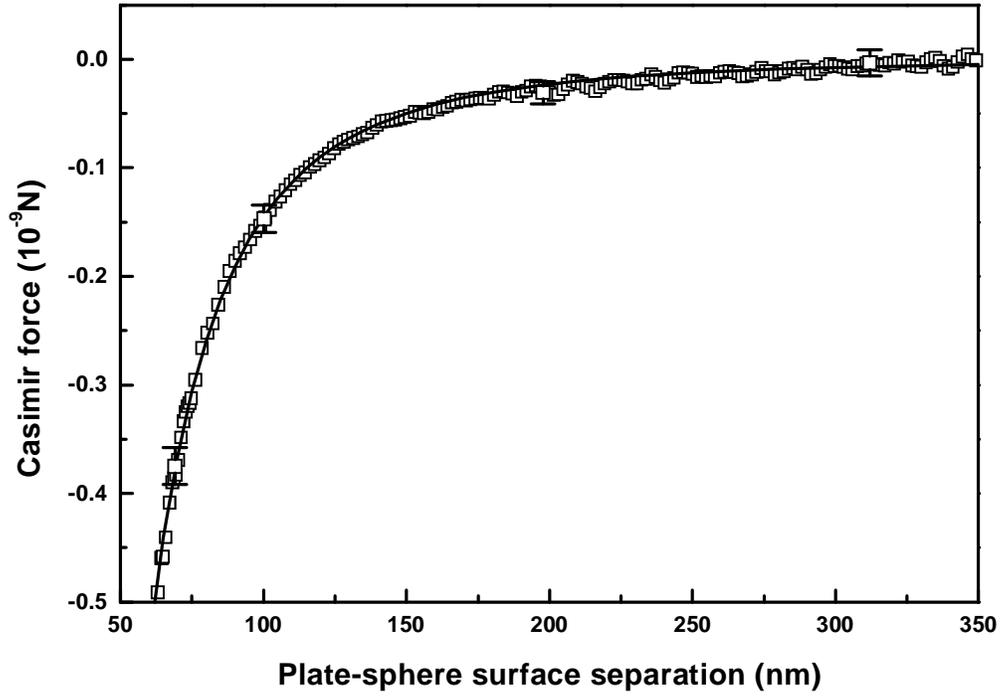